\documentclass[lettersize,journal]{IEEEtran}
\usepackage{amsmath,amsfonts}
\usepackage{algorithmic}
\usepackage{algorithm}
\usepackage{array}
\usepackage{textcomp}
\usepackage{stfloats}
\usepackage{url}
\usepackage{verbatim}
\usepackage{graphicx}
\usepackage{cite}
\usepackage{caption,subcaption}
\usepackage{hyperref}
\usepackage{bbm}
\usepackage{dsfont}
\usepackage{diagbox}
\DeclareMathOperator*{\st}{s.t.}

\begin{document}
	
	\title{\LARGE Straggler-Resilient Federated Learning over A Hybrid Conventional and Pinching Antenna Network}
	
	\author{Bibo~Wu,
		Fang~Fang,~\IEEEmembership{Senior Member, IEEE,}
		Ming Zeng,~\IEEEmembership{Member, IEEE,}
		and Xianbin Wang,~\IEEEmembership{Fellow, IEEE}
		
		\thanks{Bibo Wu, Fang Fang and Xianbin Wang are with the Department of Electrical and Computer Engineering, and Fang Fang is also with the Department of Computer Science, Western University, London, ON N6A 3K7, Canada (e-mail: {bwu293, fang.fang, xianbin.wang}@uwo.ca).}
		\thanks{Ming Zeng is with the Department of Electrical Engineering and
		Computer Engineering, Université Laval, Quebec City, QC G1V 0A6, Canada
		(e-mail: ming.zeng@gel.ulaval.ca)}}
	
	\IEEEaftertitletext{\vspace{-2.5em}}
	\maketitle
	
	\begin{abstract}
		Leveraging pinching antennas in wireless network enabled federated learning (FL) can effectively mitigate the common ``straggler" issue in FL by dynamically establishing strong line-of-sight (LoS) links on demand.
		This letter proposes a hybrid conventional and pinching antenna network (HCPAN) to significantly improve communication efficiency in the non-orthogonal multiple access (NOMA)-enabled FL system.
		Within this framework, a fuzzy logic-based client classification scheme is first proposed to effectively balance clients’ data contributions and communication conditions.
		Given this classification, we formulate a total time minimization problem to jointly optimize pinching antenna placement and resource allocation.
		Due to the complexity of variable coupling and non-convexity, a deep reinforcement learning (DRL)-based algorithm is developed to effectively address this problem.
		Simulation results validate the superiority of the proposed scheme in enhancing FL performance via the optimized deployment of pinching antenna.
		
		
	\end{abstract}
	
	
	\section{Introduction}
	\IEEEPARstart{W}{ith} the rapid proliferation of mobile edge devices and smart applications, federated learning (FL) has emerged as a promising edge artificial intelligence (AI) paradigm, enabling distributed clients to collaboratively train a global model by only exchanging the model parameters. \cite{FL_Mag}.
	This decentralized collaboration reduces communication overhead and protects the data privacy by keeping raw data local.
	However, FL systems enabled by wireless networks still face significant communication challenges stemming from limited communication resources and dynamic wireless environments. 
	These challenges often lead to the ``straggler issue" in FL systems, where slow or resource-limited clients delay the completion of each round, thereby significantly degrading model convergence and overall training efficiency.
	
	
	To address the crucial communication challenges in FL systems, advanced antenna technologies—such as fluid antennas \cite{FA_FL1}, movable antennas \cite{MA_FL}, and reconfigurable intelligent surface (RIS) \cite{RIS_FL_1, FLNOMA3}—have been explored in recent studies for their ability to improve wireless link quality.
	In particular, the study in \cite{FA_FL1} integrated fluid antennas into the FL framework to enhance system efficiency and robustness.
	In \cite{MA_FL}, movable antennas were employed to dynamically adjust antenna positions in real time, thereby overcoming adverse channel conditions.
	The work in \cite{RIS_FL_1} leveraged RIS to jointly optimize surface configuration and client selection under a unified communication-learning framework to alleviate the straggler effect, while \cite{FLNOMA3} extended it to blockage scenarios.
	
	However, fluid and movable antenna technologies typically offer only limited mobility—often constrained to within a few wavelengths—thereby significantly reducing their effectiveness in scenarios characterized by large-scale path loss. 
	Moreover, the deployment of RIS introduces challenges related to double path loss and increased control complexity.
	To address these limitations, the pinching antenna, a novel technology introduced by NTT DOCOMO in 2022 \cite{NTT}, has emerged as a low-cost and scalable solution for wireless channel reconfiguration \cite{Ding_Pin}.
	In such systems, low-cost pinches can be dynamically placed along dielectric waveguides connected to the base station, enabling the formation of additional radiation points to nearby users \cite{xiePin, Pin_1, Pin_2}.
	This mechanism could facilitate the establishment of line-of-sight (LoS) links even in the presence of obstacles or blockages.
	Thus, pinching antenna offers strong potential for improving wireless connectivity, particularly in environments with challenging propagation conditions.
	
	Conventional antennas offer stable coverage but lack adaptability for high-value yet weak-link clients, while pinching antennas provide flexible channel enhancement but face limited coverage and complex beam design. These limitations call for a communication-efficient framework that can flexibly exploit diverse wireless conditions in FL.
	As the first work to integrate pinching antenna technology into the FL framework, we propose a system of FL operating over a hybrid conventional and pinching antenna network (HCPAN), which effectively balances client resources and communication quality. 
	To further enhance communication efficiency, non-orthogonal multiple access (NOMA) transmission is employed in the proposed network.
	Specifically, a fuzzy logic–based classification scheme is devised to categorize clients into three different types.
	Based on this classification, a joint optimization problem is formulated to determine the pinching antenna placement and resource allocation, which is effectively solved using a deep reinforcement learning (DRL) algorithm.
	Simulation results demonstrate the superiority of the proposed HCPAN in enhancing overall FL performance.

	\section{System Model}

\begin{figure}
		\centering
		\includegraphics[width=3.2in]{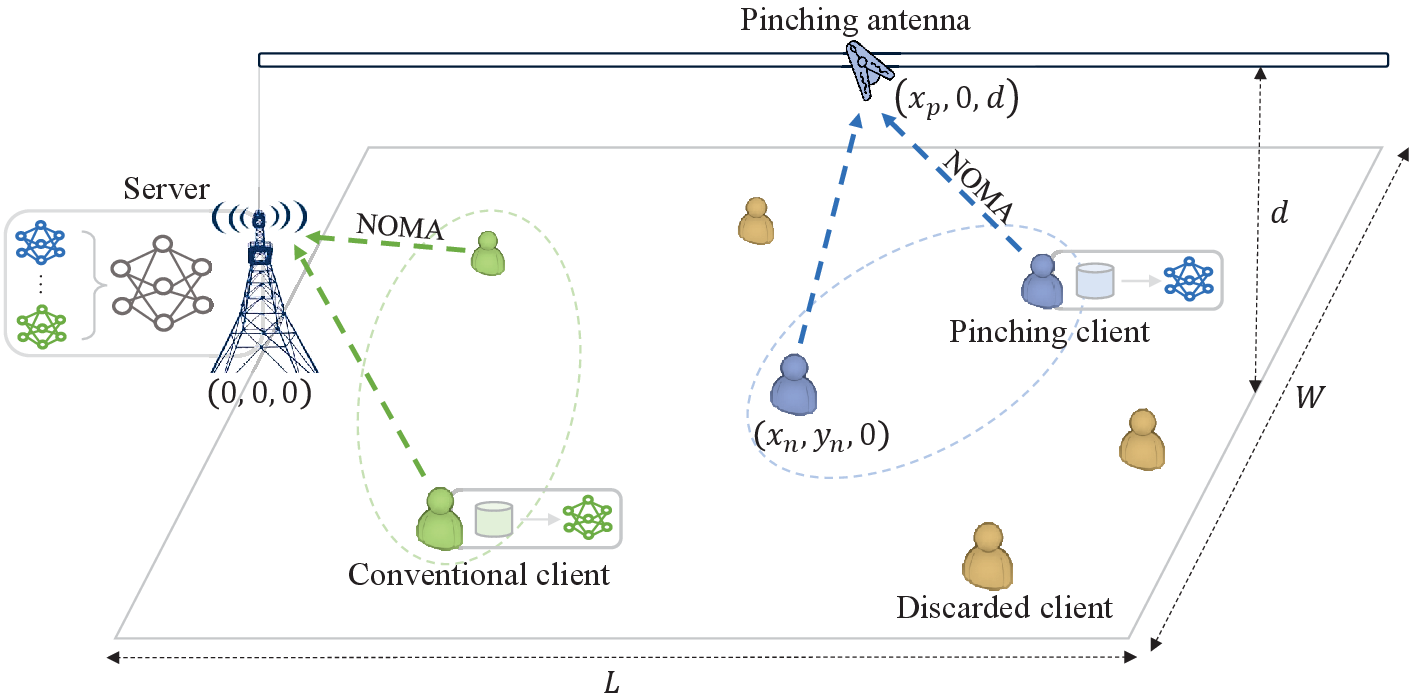}
		\caption{FL over NOMA-enabled HCPAN.}
		\label{SystemModel}
	\end{figure}

	We consider a scenario of FL involving communication-constrained stragglers.
	As illustrated in Fig. \ref{SystemModel}, a set $\cal M$ of $M$ single-antenna clients are randomly distributed within a rectangular area of dimensions $L \times W$, and their positions are denoted by ${\bf{\Psi }}_m = \left(x_m,y_m,0\right)$, $\forall m \in \cal M$. 
	The edge server is located at the center of the left boundary of the area, with position ${\bf{\Psi }}_0 = \left(0,0,0\right) $.
	To mitigate the straggler effect caused by unreliable communication conditions, we consider a pinching antenna deployed on a waveguide connected to the server at a height $d$.
	The location of the pinching antenna is denoted as ${\bf{\Psi }}_p = \left(x_p,0,d\right) $.
	
	Due to limited communication resources, only a subset $\cal N$ of $N$ clients ($N<M$) can participate in global model training during each FL round. 
	To fully exploit the diverse resources of clients, the proposed HCPAN  classifies them into three categories: conventional clients, pinching clients, and discarded clients.
	Specifically, conventional clients transmit their local model parameters to the server through the conventional antenna, while pinching clients utilize the waveguide-connected pinching antenna for model uploading. 
	Discarded clients are those unable to participate in the current round.
	
	\subsection{FL Training}
	For each selected client $n$, local model training is performed using its own labeled dataset ${{\cal{D}} _{n}} = \left\{ {\left( {{{\bf{x}}_i},{y_i}} \right)} \right\}_{i = 1}^{ {{D_{n}}} }$, where $D_{n}$ denotes the dataset size of client $n$, and ${{\bf{x}}_i}$ is the $i$-th data sample and $y_i$ is the corresponding label.
	At the $t$-th FL round, the local loss function with respect to the model parameter ${\bf{w}}_n^t$ is defined as ${{l_n}\left( {{\bf{w}}_n^t,{{\bf{x}}_i},{y_i}} \right)}$.
	Accordingly, the total loss for client $n$ is given by:
	\begin{equation}\label{}
		\begin{aligned}
			{L_n}\left( {\bf{w}}_n^t \right) = \frac{1} {{{D_n}}} \sum\limits_{i = 1}^{{D_n}} {{l_n}\left( {{\bf{w}}_n^t,{{\bf{x}}_i},{y_i}} \right)}.
		\end{aligned}
	\end{equation}
	We consider a stochastic gradient descent (SGD)-based local model update for client $n$ at the $t$-th FL round, which is
	\begin{equation}\label{}
		\begin{aligned}
			{\bf{w}}_n^t = {\bf{w}}_n^{t-1} - \alpha \nabla {L_n}\left( {{\bf{w}}_n^{t - 1}} \right),
		\end{aligned}
	\end{equation}
	where $\alpha$ denotes the learning rate.
	
	After completing local model training, each selected client uploads its trained model parameters to the server via either the conventional or pinching antenna link, depending on its assigned category. 
	For ease of notation, we define the set of $K$ conventional clients as $n \in \left\{ {1,2, \ldots ,K} \right\}$, and the set of $N-K$ pinching clients as $n \in \left\{ {K+1,K+2, \ldots ,N} \right\}$.
	The server then aggregates the received local model parameters to update the global model using a weighted averaging approach:
	\begin{equation}\label{}
		\begin{aligned}
			{{\bf{w}}^{t + 1}} = \frac{{\sum\limits_{n = 1}^N {{D_n}{\bf{w}}_n^t} }}{D} = \frac{{\left( {\sum\limits_{n = 1}^K {{D_n}{\bf{w}}_n^t}  + \sum\limits_{n = K + 1}^N {{D_n}{\bf{w}}_n^t} } \right)}}{D},
		\end{aligned}
	\end{equation}
	where $D = \sum \nolimits_{n = 1}^N D_n$ represents the total data size of all selected clients.
	Subsequently, the server broadcasts the updated global model to all clients through the conventional antenna and proceeds to the next round of FL.
	
	In this work, we reasonably neglect the downlink transmission of the global model, following the commonly adopted assumption that the server possesses sufficient transmission power and bandwidth to handle downlink communication efficiently. 
	However, it is worth noting that exploring pinching antenna–assisted downlink transmission in FL systems could be a promising direction, particularly in scenarios where physical blockages exist between the server and clients.
	
	\subsection{Computation Model}
	Let $c_n$ denote the number of CPU cycles required by client $n$ to process a single data sample.
	Assuming each client performs a one-step local update, the local training latency at client $n$ is given by
	\begin{equation}\label{}
		\begin{aligned}
			t_n^{\text{cmp}} = \frac{{{c_n}{D_n}}}{{{f_n}}},
		\end{aligned}
	\end{equation}
	where $f_n$ denotes the computational frequency of client $n$.
	The energy consumption of client $n$ for local model training is calculated as \cite{LocComEn}
	\begin{equation}\label{}
		\begin{aligned}
			e_n^{\text{cmp}} = \frac{\tau }{2}{c_n}{D_n}f_n^2,
		\end{aligned}
	\end{equation}
	where ${\tau / 2}$ represents the effective capacitance coefficient determined by the client's hardware chipset.
	
	\subsection{NOMA-based Communication Model}
	To improve the efficiency of uploading local model parameters, NOMA is employed for simultaneous uplink transmissions, while orthogonal frequency division multiple access (OFDMA) is adopted to separate the conventional-antenna and pinching-antenna links to avoid mutual interference. 
	With NOMA, the server employs successive interference cancellation (SIC) to decode the superimposed signals received from multiple clients. A channel gain–based decoding order is assumed in this work, i.e., the  receiver first decodes the signal of the client with the strongest channel gain. 
	
	For the conventional antenna-based NOMA case, the channel gains between the server and clients are ordered as ${\left| {{h_{1}^s}} \right|^2} \ge {\left| {{h_{2}^s}} \right|^2} \ge  \cdots  \ge {\left| {{h_{{K}}^s}} \right|^2}$.
	Let $p_n$ denote the transmission power of client $n$.
	Then, the achievable transmission rate from client $n$ to the server is
	\begin{equation}\label{}
		\begin{aligned}
			{R_n^{\text {con}}} = B{\log _2}\left( {1 + \frac{{{p_n}{{\left| {{h_n^s}} \right|}^2}}}{{\sum\limits_{j = n + 1}^K {{p_j}{{\left| {{h_j^s}} \right|}^2}}  + \sigma^2}}} \right),
		\end{aligned}
	\end{equation}
	where $B$ is the channel bandwidth and $\sigma^2$ is the power of additive white Gaussian noise (AWGN).
	Let $d_n$ denote the size of the local update (in bits) for client $n$, which is assumed to be identical across all clients due to the uniform dimensionality of local models.
	Based on the transmission rate ${R_n^{\text {cel}}}$, the transmission latency and energy consumption for conventional client $n$ can be expressed as:
	\begin{equation}\label{}
		\begin{aligned}
			t_n^{\text{com-con}} = \frac{{{d_n}}}{{{R_n^{\text {con}}}}}, \quad e_n^{\text{com-con}} = {p_n}{t_n^{\text{com-con}}}.
		\end{aligned}
	\end{equation} 
	
	For the pinching antenna-based NOMA case, let $c$ and $f_c$ denote the speed of light and the carrier frequency, respectively. 
	The complex channel coefficient between client $n$ and the pinching antenna is expressed as \cite{Ding_Pin}:
	\begin{equation}\label{}
		\begin{aligned}
			h_n^p = \frac{{\eta {e^{ - j\frac{{2\pi }}{\lambda }}}\left\| {{{\bf{\Psi }}_n} - {{\bf{\Psi }}_p}} \right\|}}{{\left\| {{{\bf{\Psi }}_n} - {{\bf{\Psi }}_p}} \right\|}},
		\end{aligned}
	\end{equation}
	where $\eta  = \frac{c}{{4\pi {f_c}}}$ is the path loss constant and $\lambda = \frac{c}{{{f_c}}} $ denotes the free-space wavelength.
	Note that signal attenuation within the waveguide is ignored, as the signal fading in free space is significantly greater and thus dominates the overall path loss.
	We also assume that the channels of the pinching clients are ordered as ${\left| {{h_{K+1}^p}} \right|^2} \ge {\left| {{h_{K+2}^p}} \right|^2} \ge  \cdots  \ge {\left| {{h_{{N}}^p}} \right|^2}$.
	Based on this decoding order, the achievable uplink transmission rate from client $n$ to the pinching antenna is given by:
	\begin{equation}\label{}
		\begin{aligned}
			{R_n^{\text {pin}}} = B{\log _2}\left( {1 + \frac{{{p_n}{{\left| {{h_n^p}} \right|}^2}}}{{\sum\limits_{j = n + K+1}^N {{p_j}{{\left| {{h_j^p}} \right|}^2}}  + \sigma^2}}} \right).
		\end{aligned}
	\end{equation}
	Accordingly, the transmission latency and energy consumption for client $n$ through the pinching antenna are calculated as:
	\begin{equation}\label{}
		\begin{aligned}
			t_n^{\text{com-pin}} = \frac{{{d_n}}}{{{R_n^{\text {pin}}}}}, \quad e_n^{\text{com-pin}} = {p_n}{t_n^{\text{com-pin}}}.
		\end{aligned}
	\end{equation} 
	
	\section{Fuzzy Logic-based Client Classification}
	In the proposed FL over HCPAN system, a key challenge lies in determining the category to which each client belongs—conventional client, pinching client, or discarded client. 
	Different from previous single-criterion or absolute logic-based multi-criteria schemes, this letter introduces a fuzzy logic–based client classification method that jointly considers channel quality and data contribution as input parameters \cite{fuzzybook}.
	These inputs are first transformed into fuzzy variables through a fuzzification process. 
	By applying a set of predefined fuzzy inference rules, the fuzzified inputs are mapped to corresponding fuzzy outputs. Finally, a defuzzification step is performed to convert the fuzzy output into a crisp value, which determines the final classification type of each client.
	The detailed design of the proposed scheme is presented below.

	\subsection{Input Variables}
	\subsubsection{Channel quality (CQ)} This input represents the quality of the wireless communication environment between the server and each client. 
	It plays a critical role in determining the communication latency within the FL system, as clients with poor conventional antenna channel conditions are more likely to become stragglers during model training. 
	Accordingly, in the proposed classification scheme, clients with poorer CQ are assigned a higher probability of being categorized as pinching clients or discarded clients.
	
	\subsubsection{Data contribution (DC)} This input reflects the extent to which a client can contribute to the global model update based on its local data. 
	The rationale behind this metric lies in the observation that a greater volume of local data samples can significantly improve the global FL model’s convergence, particularly in scenarios with imbalanced data distributions.
	However, the relationship between data size and model contribution is not linear; rather, it follows an increasing concave trend.
	To capture this behavior, we adopt a Weibull-based model to characterize DC of client $n$, defined as:
	\begin{equation}\label{}
		\begin{aligned}
			D{C_n} = \varpi _n - \tau  _n\exp \left( { - \lambda  _n {{D_n}  } } \right),
		\end{aligned}
	\end{equation}
	where $\varpi _n$, $\tau  _n$ and $\lambda  _n$ are predefined parameters.
	This expression reflects that while DC increases with larger local datasets, the rate of increase gradually diminishes. Therefore, it is not reasonable to classify clients solely based on their local data sizes, since a large dataset may also introduce higher computation and communication costs during FL training.
	\begin{figure}
		\centering
		\includegraphics[width=3.4in]{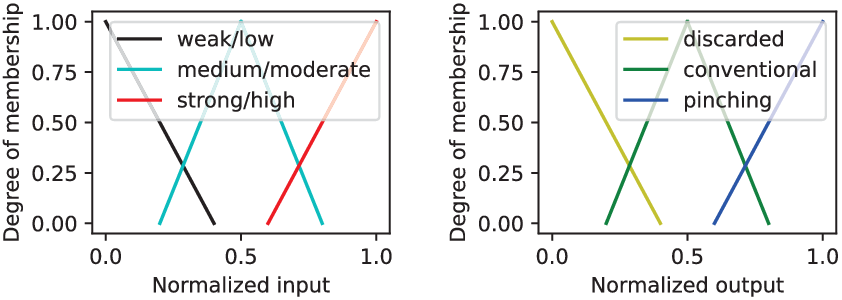}
		\caption{Degree of membership functions.}
		\label{Fuzzy_functions}
	\end{figure}
	
	\subsection{Fuzzification} 
	The two numerical inputs described above are first normalized by dividing each by its respective maximum value. 
	These normalized values are then transformed into fuzzy inputs using the predefined membership functions, as illustrated in Fig. \ref{Fuzzy_functions}. 
	The corresponding fuzzy input sets are defined as:
	\begin{itemize}
		\item CQ $\in \left\{ \text {weak, medium, strong} \right\}$,
		\item DC $\in \left\{ \text {low, moderate, high} \right\}$.
	\end{itemize}
	
	Given the two fuzzy inputs, a total of $9$ fuzzy inference rules are established, as shown in Table. \ref{fuzzy}.
	These rules establish how the fuzzy input space is related to the fuzzy output space. 
	Specifically, clients with high DC are classified as pinching clients, since the pinching antenna can offer a LoS link that ensures reliable transmission for clients with poor conventional antenna channels.
	In this case, their CQ to the server becomes less critical.
	Clients with strong CQ are assigned to the conventional category, as their superior communication conditions guarantee low transmission latency regardless of whether their DC is low or moderate. Additionally, clients with medium CQ and moderate DC are also classified as conventional clients, balancing communication efficiency and contribution.
	The remaining clients—those with both weak CQ and low or moderate DC—are categorized as discarded clients due to their limited utility and potential for straggling.
	The fuzzy output set is defined as:
	\begin{itemize}
		\item Output $\in \left\{ \text{conventional, pinching, discarded} \right\}$.
	\end{itemize}
	
	
	\begin{table}[t]
		\footnotesize
		\begin{center}
			\caption{\protect\\\textsc{Fuzzy inference rules}}
			\label{fuzzy}
			\begin{tabular}{c|c|c|c}
				\hline
				\diagbox{\textbf{DC}}{\textbf{CQ}} & \textbf{weak} & \textbf{medium} & \textbf{strong} \\
				\hline
				\textbf{low}      & discarded & discarded & conventional \\
				\textbf{moderate} & discarded & conventional  & conventional \\
				\textbf{high}     & pinching  & pinching  & pinching \\
				\hline
			\end{tabular}
		\end{center}
	\end{table}
	
	The Max–Min inference mechanism is adopted for defuzzification. 
	Specifically, the activation degree of each fuzzy rule is determined by the minimum membership value among its corresponding input variables. In cases where multiple fuzzy rules lead to the same fuzzy output, the rule with the maximum activation degree is selected to represent that output.
	To defuzzify the multiple fuzzy outputs—each corresponding to a triggered fuzzy rule—the center of gravity (COG) method is employed. 
	The defuzzified output is calculated as:
	\begin{equation}\label{COG}
		\begin{aligned}
			N{O^ * } = \frac{{\sum {NO \times f\left( {NO} \right)} }}{{\sum {f\left( {NO} \right)} }},
		\end{aligned}
	\end{equation}
	wherein $NO$ represents the normalized output level, $f\left( {NO} \right)$ is the corresponding value of the membership function, and $N{O^ * }$ is the final normalized crisp output, which determines the category to which the client is assigned.
	
	After classifying the $M$ clients into three categories using the proposed fuzzy logic scheme, clients within each category are further ranked based on their final normalized output values.
	The top $K$ conventional clients and the top $N-K$ pinching clients are selected for FL process via the conventional and pinching antenna links, respectively. 
	The remaining clients are categorized as discarded clients and excluded from the current FL round; however, they may be reconsidered in future rounds if their CQ improves. This dynamic reassessment highlights the adaptive nature of the proposed fuzzy logic–based client classification scheme.
	Note that in certain cases, the number of eligible conventional or pinching clients may fall short of $K$ or $N-K$.
	In such scenarios, discarded clients may serve as alternatives based on their normalized output values, allowing the system to maintain an adequate number of participating clients.

	\section{Problem Formulation and Solution}
	\subsection{Problem Formulation}
	Given the selected conventional and pinching clients, this section focuses on minimizing the total training latency of a single FL round through jointly optimizing the pinching antenna location and resource allocation. The optimization objective is formulated as:
	\begin{subequations}\label{OP}
		\begin{align}
			\mathop {\min }\limits_{\left\{ {x_p},{\bf{p}},{\bf{f}}\right\}} \quad & \max \left\{ {t_n^{\text {cmp}} + \tilde t_n^{\text {com}}} \right\}  \label{OP_fun} \\
			\st\ \quad  & e_n^{\text {cmp}} + \tilde e_n^{\text {com}} \le E_n^{\max }, \forall n \in {\cal N}, \label{OP_con1}\\
			& 0  \le  {p_{n}} \le p_{n}^{\text {max} }, \forall n \in {\cal N},  \label{OP_con2}\\
			& 0  \le {f_{n}} \le f_{n}^{\text {max} }, \forall n \in {\cal N}, \label{OP_con3}\\
			& 0 \le {x_p} \le L, \label{OP_con4}
		\end{align}
	\end{subequations}
	where $\bf{p} = \left\{p_n\right\}$ and $\bf{f} = \left\{f_n\right\}$ represent the collections of transmission power and computational frequency variables for all selected clients, respectively.
	The generalized variables $\tilde t_n^{\text {com}}$ and $\tilde e_n^{\text {com}}$ are defined as:
	\begin{itemize}
		\item If $n \in \left\{ 1, 2, \ldots , K\right\}$ (conventional clients): $\tilde t_n^{\text {com}} = t_n^{\text{com-con}}$, $\tilde e_n^{\text {com}} = e_n^{\text{com-con}}$; 
		\item If $n \in \left\{ K+1, K+2, \ldots , N\right\}$ (pinching clients): $\tilde t_n^{\text {com}} = t_n^{\text{com-pin}}$, $\tilde e_n^{\text {com}} = e_n^{\text{com-pin}}$.
	\end{itemize}
	Constraint \eqref{OP_con1} ensures that the total energy consumption of each selected client not exceeds its maximum allowable energy budget $E_n^{\max }$.
	Constraints \eqref{OP_con2} and \eqref{OP_con3} specify that the transmission power and computational frequency of each client are within their respective maximum limits, $p_{n}^{\text{max}}$ and $f_{n}^{\text{max}}$, respectively.
	Constraint \eqref{OP_con4} restricts the horizontal deployment range of the pinching antenna.
	
	Solving problem \eqref{OP} is challenging because of the presence of the max operator in the objective function, as well as the coupling between optimization variables in both the objective and constraint \eqref{OP_con1}. 
	These complexities make the problem non-convex and analytically intractable.
	To effectively address this issue, we resort to the DRL approach, which is capable of handling such complex, non-convex optimization problems. 
	
	\subsection{DDPG-based Solution}
	Since all optimization variables in problem \eqref{OP} are continuous, the deep deterministic policy gradient (DDPG) algorithm is well-suited for solving it. 
	To apply DDPG, we first model the problem as a Markov decision process (MDP), represented by the tuple $\left\{{S, A, R, U} \right\}$, where $S, A, R, U$ denote the state space, action space, reward function, and policy, respectively.
	
	\subsubsection{Parameter setting}
	At time slot $j$, the agent's state space is defined as ${S_j} = \left\{ D_n, {\tilde h_{n}^j, \forall n \in {{\cal N}}} \right\}$, where $D_n$ represents the local dataset size of client $n$, and $\tilde h_{n}$ denotes the generalized channel gain, covering both conventional and pinching clients.
	The action space is defined based on the optimization variables in problem \eqref{OP}, given by: ${A_j} = \left\{ {x_p^j, p_n^j,f_n^j, \forall n \in {{\cal N}}} \right\}$.
	To account for the energy budget constraint \eqref{OP_con1}, the reward function is defined as:
	\begin{equation}\label{eq_reward}
		\begin{aligned}
			R =  - {\xi _1}T + {\xi _2}\sum\limits_{n = 1}^N {{\mathop{\rm sgn}} \left( {E_n^{\max } - e_n^{\text {cmp}} - \tilde e_n^{\text {com}}} \right)},
		\end{aligned}
	\end{equation}
	where ${\xi _1}$ and ${\xi _2}$ are weighing factors, $T=\max \left\{ {t_n^{\text {cmp}} + \tilde t_n^{\text {com}}} \right\}$ represents the total latency, and $\rm sgn\left( x \right)$ is the sign function.
	
	\subsubsection{DDPG training}
	The DDPG training process is based on an actor–critic framework, which includes both target and online networks to ensure stable learning. The target actor network, denoted by $\nu \left( {S|{{\bf{\theta }}^\nu }} \right)$, with parameter ${{\bf{\theta }}^\nu }$, is in charge of action selection given the current state. 
	The target critic network, $Q\left( {S,A|{{\bf{\theta }}^Q}} \right)$, evaluates the expected value of a given state–action pair.
	Correspondingly, the online actor and online critic networks are defined as $\nu '\left( {S|{{\bf{\theta }}^{\nu '}}} \right)$ and $Q'\left( {S,A|{{\bf{\theta }}^{Q'}}} \right)$, respectively, sharing similar structures with their target counterparts.
	These online networks are updated directly during training, while the target networks are updated more slowly to stabilize learning.
	
	To improve training efficiency, the experience replay mechanism is employed.
	At each time slot $j$, the agent stores the transition tuple $\left( {{S_j},{A_j},{R_j},{S_{j + 1}}} \right)$ into a replay buffer. Mini-batches of stored experiences are then randomly sampled from this buffer to update the DDPG networks through gradient-based optimization.
	The following loss function is minimized to update the online critic network \cite{DDPG}:
	\begin{equation}\label{critic_net}
		\begin{aligned}
			L\left( {{{\bf{\theta }}^Q}} \right) = \frac{1}{{Z}}\sum\limits_j { {{{\left( {{y_j} - Q\left( {{S_j},{A_j}|{{\bf{\theta }}^Q}} \right)} \right)}^2}} },
		\end{aligned}
	\end{equation}
	where the target value ${y_j}$ is defined as ${y_j} = {R_j} + \psi Q'\left( {{S_{j + 1}},\nu '\left( {{S_{j + 1}}|{{\bf{\theta }}^{\nu '}}} \right)|{{\bf{\theta }}^{Q'}}} \right)$, $Z$ is the mini-batch size of sampled experiences and $\psi \in \left(0, 1 \right) $ denotes the discount factor.
	The online actor network is optimized using the sampled policy gradient:
	\begin{equation}\label{actor_net}
		\begin{aligned}
			{\nabla _{{{\bf{\theta }}^\nu }}}Y = {{\mathbb{E}}_{{{\bf{\theta }}^\nu }}}\left[ {Q\left( {{S_j},\nu \left( {{S_j}|{{\bf{\theta }}^\nu }} \right)|{{\bf{\theta }}^Q}} \right)} \right],
		\end{aligned}
	\end{equation}
	where ${\mathbb{E}}\left(  \cdot  \right)$ denotes the expectation operator.
	To ensure training stability, the following soft update rule is utilized to update target actor and critic networks:
	\begin{equation}\label{target_update}
		\begin{aligned}
			{{\bf{\theta }}^{\nu '}} \leftarrow \zeta {{\bf{\theta }}^\nu } + \left( {1 - \zeta } \right){{\bf{\theta }}^{\nu '}}, \quad {{\bf{\theta }}^{Q'}} \leftarrow \zeta {{\bf{\theta }}^Q} + \left( {1 - \zeta } \right){{\bf{\theta }}^{Q'}},
		\end{aligned}
	\end{equation}
	where $\zeta$ indicates the updating rate.

	\section{Simulation Analysis}
	This section verifies the performance of the proposed FL over HCPAN through simulation results.
	We consider a rectangular area of dimensions $30$ m $\times 10$ m, where the server is positioned at the midpoint of the left boundary, and $M = 30$ clients are randomly distributed within the service region.
	We set the height of the pinching antenna as $d = 3$ m.
	At each FL round, $N = 6$ clients are selected for global model training, with an equal split of $3$ clients utilizing conventional antenna and $3$ utilizing pinching antenna.
	The carrier frequency is set to $f_c = 3.5$ GHz, the path loss exponent is $2.4$, and the noise power spectral density is set to $-174$ dBm/Hz.
	All other parameters follow those in \cite{DT_FL_Bibo}.
	
	
	
	\begin{figure}[t]
		\centering
		\begin{subfigure}[b]{0.24\textwidth}
			\includegraphics[width=\textwidth]{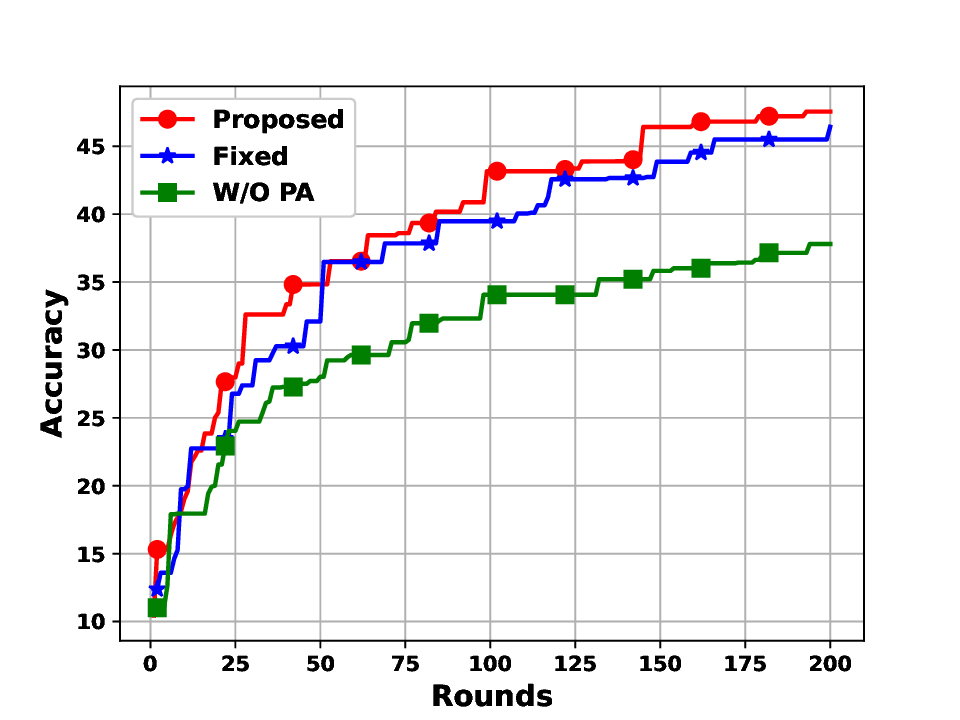}
			\caption{}
			\label{FL_acc}
		\end{subfigure}
		\begin{subfigure}[b]{0.24\textwidth}
			\includegraphics[width=\textwidth]{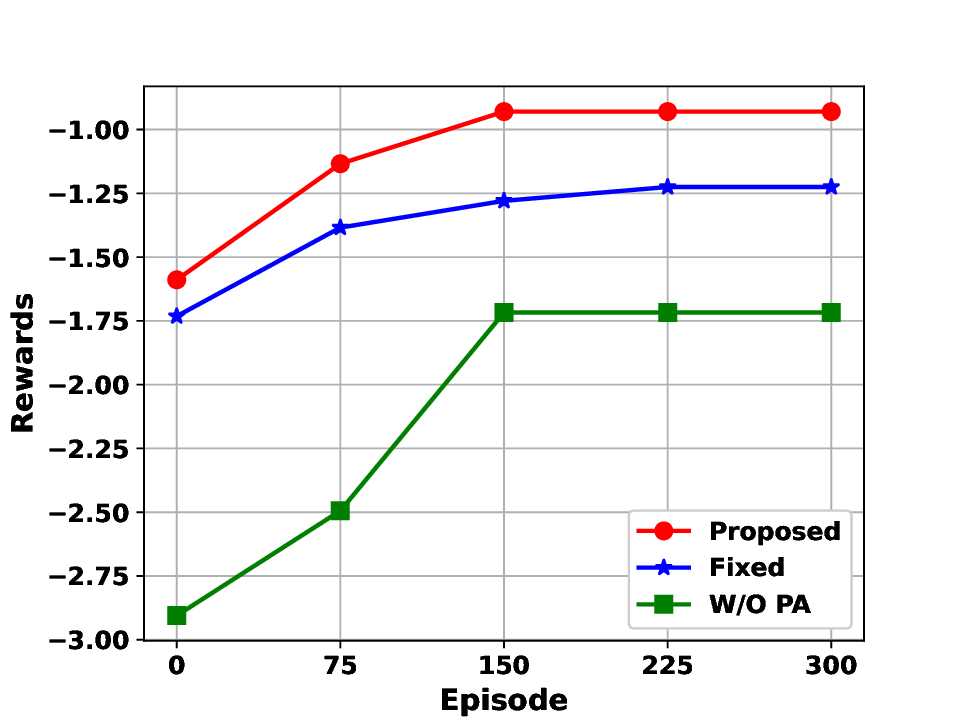}
			\caption{}
			\label{Reward}
		\end{subfigure}
		\caption{(a) FL accuracy performance; (b) Reward performance.}
		\label{FML_accuracy_mul}
	\end{figure}
	
	Fig. \ref{FML_accuracy_mul}\subref{FL_acc} illustrates the FL accuracy performance of the proposed scheme compared with two benchmark schemes: the fixed-placement pinching antenna scheme and the scheme without a pinching antenna (W/O PA) where two NOMA-based conventional antenna groups are employed for a fair comparison.
	The training is conducted on the non-independent and identically distributed (non-IID) CIFAR-10 dataset.
	To avoid visual clutter among curves, the maximum accuracy metric is adopted to present the performance, following the approach in \cite{KaidiFL}.
	We can observe that the proposed scheme achieves the highest FL accuracy across training rounds.
	This improvement stems from the hybrid deployment of conventional and pinching antennas, as well as the dynamic optimization of the pinching antenna’s location.
	In particular, clients with high data contributions—who would otherwise be stragglers in conventional antenna-only settings due to poor communication quality—are enabled to participate in global model training through optimized pinching links.
	
	Fig. \ref{FML_accuracy_mul}\subref{Reward} presents the reward performance of DDPG-based solutions under the three considered schemes.
	For fair comparison, the two benchmark schemes also adopt the DDPG algorithm for resource allocation, following the same reward definition as given in \eqref{eq_reward}.
	We can find that the proposed scheme consistently outperforms the two benchmarks during the DDPG training process and achieves the highest reward at convergence.
	The performance gain is attributed to the dynamic optimization of the pinching antenna’s location, which provides better adaptability compared to the fixed-location scheme, and to the improved channel conditions enabled by the pinching antenna compared to the W/O PA scheme.
	These results validate the effectiveness of the proposed FL framework over HCPAN.

	\section{Conclusion}
	In this letter, we proposed a communication-efficient system that integrates conventional and pinching antennas within a NOMA-enabled network. It aims to improve the convergence performance of FL by mitigating the straggler issue.
	In this novel framework, we tackled the newly introduced challenge of balancing clients' data contributions with their communication conditions. Our work also verified that both the deployment and optimized placement of the pinching antenna effectively improve FL performance.
	Potential extensions of this work include incorporating multiple waveguides and/or multiple pinching antennas along a single waveguide, enabling more stragglers to establish high-quality links and thereby further enhancing overall FL performance.
	

	\bibliographystyle{IEEEtran}
	\bibliography{EEref}
\end{document}